\documentclass{article}
\usepackage[utf8]{inputenc}
\usepackage{amsmath,amssymb}
\usepackage{mathtools}
\usepackage{multirow}
\usepackage{relsize}
\usepackage{tikz}
\usetikzlibrary{arrows,decorations.pathmorphing}
\usepackage{tkz-berge}
\usepackage{subfigure}
\usepackage{authblk}

\newcommand{\ket}[1]{| #1 \rangle}
\newcommand{\bra}[1]{\langle #1 |}
\newcommand{\ketbra}[2]{| #1 \rangle \langle #2 |}

\newcommand{\kr}{\otimes}
\newcommand{\tr}{\mathrm{Tr}}
\newcommand{\Id}{\mathrm{I}}

\newcommand{\Bn}{\mathcal{L}}
\newcommand{\HH}{\mathcal{H}}

\newcommand{\diag}{\operatorname{diag}}

\newcommand\commonTikzFragment{
	\Vertex[x=0,y=0]C
	\Vertex[x=1.5,y=2.5]B
	\Vertex[x=-1.5,y=2.5]A
	\tikzstyle{EdgeStyle}=[post]
	\Edge[](A)(C)
	\Edge[](B)(C)
}
\title{Quantum inferring acausal structures and the Monty Hall problem}
\author[$*$,$\dagger$]{Dariusz Kurzyk}
\author[$*$,$\dagger$]{Adam Glos}
\affil[$*$]{Institute of Theoretical and Applied Informatics, Polish Academy
              	of Sciences, Ba{\l}tycka 5, 44-100 Gliwice, Poland}
\affil[$\dagger$]{Institute of Mathematics, Silesian University of Technology, 
                Kaszubska 23, 44-100 Gliwice, Poland}

\date{Gliwice \today}

\begin{document}

\maketitle

\begin{abstract}
This paper presents a quantum version of the Monty Hall problem based upon the quantum inferring acausal structures, which can be identified with generalization of Bayesian networks. Considered structures are expressed in formalism of quantum information theory, where density operators are identified with quantum generalization of probability distributions. Conditional relations between quantum counterpart of random variables are described by quantum conditional operators. Presented quantum inferring structures are used to construct a model inspired by scenario of well-known Monty Hall game, where we show the differences between classical and quantum Bayesian reasoning.
\end{abstract}

\section{Introduction}
Probability theory has a lot of applications in many areas of science and 
engineering. In particular probabilistic modelling has strong impact in development 
of artificial intelligence, providing tools for knowledge representation, 
knowledge management and reasoning \cite{bishop2006pattern}. Many problems related 
to computer vision, speech recognition, extraction of information or diagnosis of 
diseases can be modelled by probabilistic graphical models which structures 
describe conditional dependencies between random variables 
\cite{koller2009probabilistic}. This kind of models is indirectly related with 
Bayesian reasoning, which is a method of statistical inference based on Bayes' 
theorem. Bayesian reasoning gives a possibility of decision making under uncertainty, which leads to updating of beliefs in light of new information. The Bayesian approach to reasoning can be used in the well-known Monty Hall problem, which can be considered as a counterintuitive two-step decision problem. The probabilistic analysis of this problem develops an intuition about updating a prior belief as new evidence becomes, which gives a deeper understanding of Bayesian methods used in the artificial intelligence. 
Consequently, we use the Monty Hall problem as an illustration of quantum Bayesian 
reasoning based on quantum information theory and quantum inferring acausal 
structures.

Cerf and Adami \cite{cerfadami1997,cerfadami1999} introduced a quantum conditional amplitude operator as an extension of conditional probability distribution. This approach allowed to complement of relationship between Shannon conditional entropy and von Neumann conditional entropy. In this context, quantum conditional operators are considered as a generalization of classical conditional probability distributions.

Definition of conditional states was extended in the context of quantum channel by 
Leifer \cite{leifer2006}. He connected classical conditional probabilities and 
bipartite density operators with variant of well-known Jamiołkowski isomorphism. 
Subsequently, Leifer with Spekkens \cite{leifer2013towards,leiferspekkens2014} 
presented formalism of quantum inferring structures based on quantum conditional 
states. They described two cases of structures: causal (one system at two times) 
and acausal (two systems at a single time).

Quantum inferring structures based on quantum probability theory allow for a novel causal interpretation and generalization of the results of quantum mechanics. Recently, in many scientific disciplines, the inferring causal and acausal relations from observed correlations are relevant problems. The framework of
quantum conditional operators was used by Brunker \cite{brukner2014quantum} to
investigate quantum causality. The author assumed, that indefinite causal
structures could provide methodological tools in quantum theories of gravity. 
In \cite{riedspekkens2014}, the authors used the quantum conditional operators
as  causal maps to consideration of the problem of causal inference for quantum
variables. They introduced the concept of causal tomography, which unifies
conventional quantum tomography schemes and provides a complete solution of the causal
inference problem using a quantum analogue of a randomized trial.

\section{Quantum conditional operators}
In \cite{leiferspekkens2014,leifer2013towards}, Leifer and Spekkens introduced a definition of {\it elementary region} as a quantum system at fixed point in time. By them, the relations between regions were described in two contexts. In the first case, elementary regions $A,B$ are {\it causally} related. It means that there is casual inference from $A$ to $B$, e.g. it can be a single system at two times. In the second case, the $A,B$ are {\it acausally} related, it means that elementary regions represent two distinct systems at a fixed time. In this paper, the emphasis is put on the elementary regions acausally related. 

Let $\Bn(\HH)$ be the set of linear operators on complex Hilbert space
$\HH$. The classical probability distribution can be generalized by density operator $\rho_A\in\Bn{(\HH_A)}$ ($\rho_A=\rho^\dagger$, $\rho_A\geq0$, $\tr(\rho_A)=1$). The analogue of joint probability distribution for
quantum systems $AB$ is a density operator
$\rho_{AB}\in\Bn(\HH_{A}\kr\HH_{B})$. Moreover, the analogue of marginalization over a joint distribution is the partial trace of quantum state, i.e.
$\rho_A=\tr_B(\rho_{AB})$. An {\it acausal conditional state}
\cite{leiferpoulin2008} for $B$ given $A$ is a positive operator
$\rho_{B|A}\in\Bn(\HH_{AB})$ that satisfies
\begin{equation}
	\tr_B(\rho_{B|A})=\Id_{A},
\end{equation}
where $\Id_A$ is the identity operator from $\Bn(\HH_{A})$. Conditional operators 
can be defined by the usage of $\star$ product \cite{leifer2013towards}, where 
$\star : \Bn(\HH)\to\Bn(\HH)$ and
\begin{equation}
	\sigma_A\star\sigma_B = \sigma_B^\frac12 \sigma_A \sigma_B^\frac12,
\end{equation}
for  $\sigma_A,\sigma_B\in\Bn(\HH)$.
Hence, the conditional operator is expressed as
\begin{equation}
	\rho_{B|A}=\rho_{AB}\star(\rho_A^{-1}\kr\Id_B)=(\rho_A^{-\frac{1}{2}}\otimes\Id_B)
	 \rho_{AB}(\rho_A^{-\frac{1}{2}}\otimes\Id_B),
\end{equation}
where $(\cdot)^{-1}$ denotes the Moore-Penrose pseudoinverse. Similarly as in classical probability theory, the following equation holds true
\begin{equation}\label{eq:joint}
\rho_{AB}=\rho_{B|A}\star(\rho_A\kr\Id_B).
\end{equation}
The occurrence of the event 
$\ketbra{b}{b}$ on the subsystem $B$ causes the transformation of $\rho_{AB}$ into
\begin{align}\label{eq:joint2}
	\rho_{A,B= b} &= \frac{(\Id_A\kr \ketbra{b}{b})\rho_{AB}(\Id_A\kr 
	\ketbra{b}{b})}{\tr((\Id_A\kr \ketbra{b}{b})\rho_{AB})}.
\end{align}
The comma in index of $\rho_{A,B=b}$ is used only for readability. The marginal
probability distribution acting on subsystem $A$ after the transformation is
given as
\begin{equation}\label{glosKurzyk:simplCond}
	\rho_{A}^{B= b} = \tr_B ( \rho_{A,B= b}).
\end{equation}
Summing up, given an initial $\rho_A$ one can compute an updated $\rho_{A}^{B=b}$ by first computing $\rho_{AB}$ using $\rho_{B|A}$ and Eq.~(\ref{eq:joint}) and the following Eqs.~(\ref{eq:joint2}) and (\ref{glosKurzyk:simplCond}).

%Suppose, that  $\rho\in\Ln(\C^{n_A}\kr\C^{n_B})$, then three cases of relationships between quantum variables $A$ and $B$ can be considered.
%\begin{enumerate}
%	\item \emph{Independent random variables}. If there exist $\rho_A\in\Ln(\C^{n_A})$ and $\rho_B\in \Ln(\C^{n_B})$ such that $\rho = \rho_A \kr \rho_B$, then we get analogy to independence of random variables.
%	\item \emph{Separable case}. If $\rho$ is separable, but there don't exist $\rho_A\in\Ln(\C^{n_A})$ and $\rho_B\in \Ln(\C^{n_B})$ such that $\rho = \rho_A \kr \rho_B$, then we get analogy to dependence of random variables.
%	\item \emph{Entangled case}. State $\rho$ is entangled, then there exist correlations which cannot be found in classical probability theory.
%\end{enumerate}

\section{Inferring acausal structures}
The graphical model is defined as a probabilistic model for which the conditional 
dependence between random variables are defined by a graph. It can be used for statistical inference under a set of random variables. Replacing 
classical random variables by their quantum counterparts gives a class of quantum 
structures, which can be used for inferring amongst a set of quantum states.

\begin{figure}
	\centering
	\begin{tikzpicture}[scale=0.7,transform shape]
	\Vertex[x=0,y=0]B
	\Vertex[x=0,y=3]A
	\tikzstyle{EdgeStyle}=[post]
	\Edge[](A)(B)
	\end{tikzpicture}
	\caption{Graphical interpretation of relationship between variables represented 
	by quantum states. The state of quantum system $B$ is expressed as 
	$\rho_B=\tr_A(\rho_{B|A}\star(\rho_A\kr \Id_B))$.}
	\label{glosKurzyk:AtoB}
\end{figure}
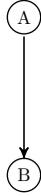

Let us consider operators $\rho_{B|A}\in\Bn(\HH_{AB})$ and $\rho_A\in\Bn(\HH_{A})$. The state $\rho_{B|A}$ characterizes conditional relationship between systems $A$ and $B$. In this case, the state of system $B$ can be expressed as
\begin{equation}\label{exp:lawoftotalprobability}
	\rho_B = \tr_A(\rho_{B|A}\star(\rho_A\kr\Id_B)).
\end{equation}
In classical probability theory, this formula is called the law of total probability. Figure~\ref{glosKurzyk:AtoB} illustrates graphical interpretation of \label{lawoftotalprobability} reasoning over  $\rho_{B|A}$ and $\rho_A$, what can be considered as simple Bayesian network.

The conditional relationship associated with graphical structure of the
network allows to rewrite the joint state of $A$ and $B$ as
follows
\begin{equation}\rho_{AB}= \rho_{B|A}\star ( \rho_A \kr \Id_B).\end{equation}

Consider the structure given in Fig.~\ref{glosKurzyk:AtoBC} described by state 
$\rho_A$ and conditional states $\rho_{B|A}$ and $\rho_{C|A}$. To obtain joint 
state of $A$ and $B$ or $A$ and $C$, we use formulas
\begin{equation}
	\begin{split}
		\rho_{AB}&= \rho_{B|A}\star ( \rho_A \kr \Id_B), \\
		\rho_{AC}&= \rho_{C|A}\star ( \rho_A \kr \Id_C).
	\end{split}
\end{equation}
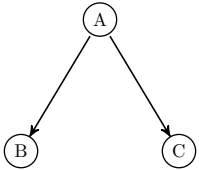
\begin{figure}
	\centering
	\begin{tikzpicture}[scale=0.7,transform shape]
	\Vertex[x=-1.5,y=0]B
	\Vertex[x=1.5,y=0]C
	\Vertex[x=0,y=2.5]A
	\tikzstyle{EdgeStyle}=[post]
	\Edge[](A)(B)
	\Edge[](A)(C)
	\end{tikzpicture}
	\caption{Inferring structure with three random variables represented by $\rho_A,\rho_{B|A}$ and $\rho_{C|A}$.}
	\label{glosKurzyk:AtoBC}
\end{figure}
The join state of $A$, $B$ and $C$ can be obtained by using the formula
\begin{equation}\label{glosKurzyk:revQuantCond}
	\rho_{ABC}= (\rho_{C|A}\kr\Id_B) \star (\rho_{B|A}\kr\Id_C) \star (\rho_{A}\kr\Id_B\kr\Id_C).
\end{equation}

Acausal relations between quantum states can be described by quantum Bayesian networks \cite{leiferpoulin2008} defined as a pair $(\mathbb{G}, \mathbb{D})$ where 
\begin{itemize}
	\item $\mathbb{G}=({\bf X}, {\bf E})$ is a directed acyclic graph, where each 
	vertex $X_i\in {\bf X}$ is associated to a quantum system with space 
	$\Bn(\HH_{i})$.
	\item $\mathbb{D}=\{\rho_{{X_i}|\Pi_{X_i}}:1\leq i \leq N\}$ is the set of
	the conditional operators between node $X_i$ and its parents $\Pi_{X_i}$.
\end{itemize}

The quantum Bayesian network can be interpreted as a quantum system associated
with space $\Bn(\HH_{1}\otimes\cdots\otimes\HH_{N})$.  The conditional
independence relationship associated with a graphical structure of Bayesian
network allows to rewrite the generalized joint distribution of ${\bf X}$ as
follows
\begin{equation}\label{eq:bn}
\rho_{X_1X_2\dots X_N}=\big(\bigstar\big)_{i=1}^N \rho_{{X_i}|\Pi_{X_i}}
\end{equation}
where $\big(\bigstar\big)_{i=1}^N 
\rho_{{X_i}|\Pi_{X_i}}=\rho_{{X_1}|\Pi_{X_1}} 
\star\cdots\star\rho_{{X_N}|\Pi_{\mathcal{X}_N}}$.
In Eq.~(\ref{eq:bn}), each factor $\rho_{{X_i}|\Pi_{X_i}}$ should be expressed 
as $\Id\kr\rho_{{X_i}|\Pi_{X_i}}\kr\Id$, but we drop identity 
operators and Kronecker products for simplicity.

Leifer and Poulin \cite{leiferpoulin2008} emphasize the fact, that quantum graphical models have significant applications in quantum error corrections and the simulation of many-body quantum systems.

\section{Acausal structure in Monty Hall game} \label{glosKurzyk:QuantProblem}

In this section, we present the usage of the quantum inferring structures in 
modelling a quantum system inspired by a well-known Monty Hall problem based on the 
popular game show {\it Let's Make a Deal}. The problem was first  described by 
Selvin in \cite{selvin1975a,selvin1975b}. Subsequently, the Monty Hall problem has 
been considered as a subject of many investigations, i.e. this issue is modelled 
through a formal application of Bayes' rule \cite{fenton2012risk,gill2014bayesian}. In \cite{li2001quantum}, the Monty Hall game was generalized to the quantum domain. The paper shows that fair zero-sum game can be realized if the quantum measurement strategy is permitted. 
A more detailed discussion about quantum Monty Hall can be found at \cite{ariano2002quantum,flitney2002quantum}, where authors consider many scenarios. In \cite{gawron2010,khan2010quantum}, the effect of decoherence on results of the game was analysed. 

Classical scenario of the Monty Hall game is as follows. 
There are two participants of the game: host and player. Assume that three doors 
labelled $0,1$ and $2$ are closed. A car is behind one of the doors, while the 
goats are behind the other two. The host knows which of the three doors hides the 
car, but the player does not know where the main prize is hidden. The player picks 
a 
door and subsequently, the host opens one of the remaining doors, revealing a goat. 
Finally, the player selects one of the remaining doors. The question is: should the 
player stick to his/her original choice or pick the other door?

\subsection{Classical case}\label{glosKurzyk:secNoEntCase}
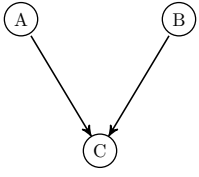
\begin{figure}
	\centering
	\begin{tikzpicture}[scale=0.7,transform shape]
	\commonTikzFragment
	\end{tikzpicture}
	\caption{Bayesian network describing Monty Hall game.}
	\label{glosKurzyk:noEntGraph}
\end{figure}

The Monty Hall game can be modelled as a simple Bayesian network with three random variables $X_A,X_B,X_C$, whose graph is presented in Fig.~\ref{glosKurzyk:noEntGraph}.
The variable $X_A$ represents the player's information about which door contains the prize. Initially, the player do not have any knowledge about this, and thus we assume that probability distribution $P(X)$ is uniform and is given by $(1/3,1/3,1/3)$. The $X_B$ describes the first choice of 
the player and we also assume that the probability distribution of $X_B$ is 
$(1/3,1/3,1/3)$. The random variable $X_C$ depends on the $X_A,X_B$ and describes  
selected door by the host. The probability distribution $P(X_C|X_A,X_B)$ is given 
in Table~\ref{glosKurzyk:tableRhoCAB}.

\begin{table}[h]
\caption{Probability of choosing door by host.}
\label{glosKurzyk:tableRhoCAB}      % Give a unique label
\centering
\begin{tabular}{|cc|cc|ccc|}
		\cline{5-7}
		\multicolumn{4}{c|}{ } & \multicolumn{3}{|c|}{Host} \\
		\multicolumn{4}{c|}{ } & No. 0 & No. 1 & No.2 \\\hline
		\multirow{9}*{Prize}& \multirow{3}*{No. 0} & \multirow{3}*{Player} 
		& No. 0 & 0 & $\frac{1}{2}$ & $\frac{1}{2}$ \\ 
		& & & No. 1 & 0 & 0 & 1 \\
		& & & No. 2 & 0 & 1 & 0 \\\cline{3-7}
		& \multirow{3}*{No. 1} & \multirow{3}*{Player} 
		& No. 0 & 0 & 0 & 1 \\ 
		& & & No. 1 & $\frac{1}{2}$ & 0 & $\frac{1}{2}$ \\
		& & & No. 2 & 1 & 0 & 0 \\\cline{3-7}
		& \multirow{3}*{No. 2} & \multirow{3}*{Player} 
		& No. 0 & 0 & 1 & 0 \\ 
		& & & No. 1 & 1 & 0 & 0 \\
		& & & No. 2 & $\frac{1}{2}$ & $\frac{1}{2}$ & 0 \\\hline
	\end{tabular}
\end{table}

%\begin{table}[ph]
%	\tbl{Comparison of acoustic for frequencies for 
%		piston-cylinder problem.}
%	{\begin{tabular}{@{}cccc@{}} \toprule
%			Piston mass & Analytical frequency & TRIA6-$S_1$ model &
%			\% Error \\
%			& (Rad/s) & (Rad/s) \\ \colrule
%			1.0\hphantom{00} & \hphantom{0}281.0 & \hphantom{0}280.81 & 0.07 \\
%			0.1\hphantom{00} & \hphantom{0}876.0 & \hphantom{0}875.74 & 0.03 \\
%			0.01\hphantom{0} & 2441.0 & 2441.0\hphantom{0} & 0.0\hphantom{0} \\
%			0.001 & 4130.0 & 4129.3\hphantom{0} & 0.16\\ \botrule
%		\end{tabular}}
%	\end{table}

Now, given the computational basis $\{\ket{ijk}\}$ for $0\leq i,j,k\leq2$, we create an analogous quantum Bayesian network, where the random distributions $P(X_A)$, $P(X_B)$, $P(X_C|X_A,X_B)$ are replaced by 
\begin{equation}\label{probs}
	\begin{split}
		\rho_A &= \diag\left (\mathsmaller{\frac{1}{3}},\mathsmaller{\frac{1}{3}},\mathsmaller{\frac{1}{3}}\right ), \\
		\rho_B &= \diag\left (\mathsmaller{\frac{1}{3}},\mathsmaller{\frac{1}{3}},\mathsmaller{\frac{1}{3}}\right ), \\
		\rho_{C|AB} &= \diag\left (0,\mathsmaller{\frac{1}{2}},\mathsmaller{\frac{1}{2}},0,0,1,0,1,0,  \right .\\
		&\phantom{\:=\diag(}\left .0,0,1,\mathsmaller{\frac{1}{2}},0,\mathsmaller{\frac{1}{2}},1,0,0, \right .\\
		&\phantom{\:=\diag(} \left . 0,1,0,1,0,0,\mathsmaller{\frac{1}{2}},\mathsmaller{\frac{1}{2}},0 \right ).
	\end{split}
\end{equation}
Next, we determine 
\begin{equation}\label{glosKurzyk:cauntingFirstFormula}
	\rho_{ABC} = \rho_{C|AB} \star ( \rho_{A}\kr \Id_B\kr \Id_C) \star ( \Id_A\kr \rho_B\kr \Id_C),
\end{equation}
which describes the state of composite quantum system $ABC$ identified with graph in Fig.~\ref{glosKurzyk:noEntGraph}. Note that $\star$ product is non-associative in general but when the factors in the product commute it is associative. Thus, taking into account given diagonal matrices $\rho_A$, $\rho_B$ and $\rho_{C|AB}$, obtained state (\ref{glosKurzyk:cauntingFirstFormula}) can be expressed in form ${\rho_{ABC} =\frac{1}{9}\rho_{C|AB}}$.  

Let us assume that the player chooses $b$-th door,  where doors are labelled by $0$, $1$ and $2$. Hence, we assume that measurement labelled by $b$ is performed on subsystem $B$ and the state $\rho_{AB}$ is transformed into
\begin{equation}
	\rho_{A,B= b,C} = \frac{(\Id_A\kr\ketbra{b}{b}\kr \Id_C) 
	\rho_{ABC}(\Id_A\kr\ketbra{b}{b}\kr \Id_C)}{\tr((\Id_A\kr\ketbra{b}{b}\kr 
	\Id_C) \rho_{ABC})}.
\end{equation}
Afterwards, the host opens the $c$-th door. Similarly, the measurement labelled by $c$ is performed on subsystem $C$ and we get
\begin{equation}
	\rho_{A,B= b,C= c} = \frac{(\Id_A\kr \Id_B \kr \ketbra{c}{c}) \rho_{A,B= 
	b,C}(\Id_A\kr \Id_B \kr \ketbra{c}{c})}{\tr((\Id_A\kr \Id_B \kr \ketbra{c}{c}) 
	\rho_{A,B= b,C})}.
\end{equation}
After performed measurements,  state of subsystem $A$  is given by
\begin{equation}
	\rho_A^{B= b,C= c} = \tr_{BC}( \rho_{A,B= b,C= c}). \label{glosKurzyk:cauntingLastFormula}
\end{equation}
State $\rho_A$ has encoded a priori information about position of prize, which is updated under observed evidences. Thus, state $\rho_A^{B= b,C= c}$ represents updated player's knowledge about doors which can contain the prize, with assumption that firstly the player chose door $b$ and the host opened door $c$.

Since, the host knows where the car is hidden, we assume that $b \not = c$.
Suppose, that player chooses 0-th door and the host opens $1$-st door. Hence, using Eqs. (\ref{glosKurzyk:cauntingFirstFormula})-(\ref{glosKurzyk:cauntingLastFormula}) we get
\begin{equation}
	\label{eq:sol01}
	\rho_A^{B= 0,C= 1}=\begin{bmatrix}
		\frac{1}{3} & 0 & 0 \\
		0 & 0 & 0 \\
		0 & 0 & \frac{2}{3}
	\end{bmatrix}.\end{equation}
The solution obtained above corresponds to optimal player's strategy in classical 
Monty Hall game. According to the probability distribution (\ref{eq:sol01}), the 
player should change his/her first decision. In general, for any 
$a,b,c\in\{0,1,2\}$ such that $a \not = b$, $b \not = c$ and $a \not = c$  equality 
$\bra{a} \rho_A^{B= b,C= c} \ket{a}=\frac23$ holds. It means that, if the player 
changes his/her first choice, then the probability of winning the car is  $\frac23$ 
.

\subsection{Non-classical case}\label{glosKurzyk:secEntCase} In the non-classical
case, we  assume, that systems $A$ describing the prize position and $B$ describing the first choice of the player are entangled. The state $\rho_{AB}$ must
satisfy following conditions:
\begin{enumerate}
\item $\bra{k}\tr_A (\rho_{AB})\ket{k} = \bra{k}\tr_B (\rho_{AB})\ket{k} = \frac{1}{3}$ for $k\in\{0,1,2\}$,
\item $\bra {c}\rho_{A}^{B=b,C=c} \ket{c}=0$ for all $b,c\in\{0,1,2\}$ such that $b\neq c$.
\end{enumerate}
The second condition implies, that the host cannot open the door with the prize
behind it. In our considerations, the behaviour of player and host is similar like in classical case. It means that the player do not have access to quantum strategies. However, we assume that quantum effects occur between position of prize and player's first decision  but participants of the game don't know about it.

\begin{figure}
	\centering
	\begin{tikzpicture}[scale=0.7,transform shape]
	\commonTikzFragment;
	\draw[decorate,decoration=zigzag] (A) -- (B);
	\end{tikzpicture}
	\caption{Zigzag line means quantum dependence between states of systems $A$ and $B$ (entanglement) and arrows mean classical dependencies.}
	\label{glosKurzyk:entGraph}
\end{figure}
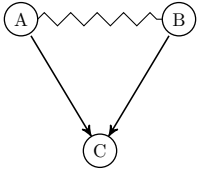

Suppose that system is in state 
\begin{equation}
\tilde\rho_{AB} = \frac{1}{3} ( \ket{00} + \ket{11} + \ket{22})( \bra{00} + 
\bra{11} + \bra{22}).
\end{equation}
After tracing-out this state we get
\begin{equation}
\begin{split}
\tilde\rho_A &= \tr_B (\tilde \rho_{AB} ) =  \diag \left ( \mathsmaller{\frac{1}{3}},\mathsmaller{\frac{1}{3}},\mathsmaller{\frac{1}{3}}\right ), \\ 
\tilde\rho_B &= \tr_A (\tilde \rho_{AB} ) = \diag \left ( \mathsmaller{\frac{1}{3}},\mathsmaller{\frac{1}{3}},\mathsmaller{\frac{1}{3}} \right ).
\end{split}
\end{equation}
Thus $\tilde\rho_A$ and $\tilde\rho_B$ are maximally mixed and given reduced states correspond to the states from the  previous section. Since states of systems $A$ and $B$ are entangled, we describe
a inferring structure of this scenario according to
Fig.~\ref{glosKurzyk:entGraph}.

Let us notice, that $\tilde\rho_{AB}$ is a pure state, hence the equations  $\tilde{\rho}_{AB}^{\frac{1}{2}}=\tilde{\rho}_{AB}$ and $\tilde{\rho}_{ABC}=\rho_{C|AB}\star(\tilde\rho_{AB}\kr\Id_C)=(\tilde\rho_{AB}\kr\Id_C)\rho_{C|AB}(\tilde\rho_{AB}\kr\Id_C)$ hold true. Using the same state $\rho_{C|AB}$ as in (\ref{probs}) and proceeding similarly like in (\ref{glosKurzyk:cauntingFirstFormula})--(\ref{eq:sol01}) we get for $b=0$ and $c=1$ following results 
\begin{equation}
	 \tilde\rho_A^{B= 0,C= 1} = \begin{bmatrix}
		1 & 0 & 0\\
		0 & 0 & 0\\
		0 & 0 & 0
	\end{bmatrix}, \\
\end{equation}
Note that $\bra{0}\tilde\rho_A^{B= 0,C= 1}\ket{0}=1$. In general, the equality
$\bra{b} \tilde\rho_A^{B= b,C= c} \ket{b}=1$ holds for any $b,c\in\{0,1,2\}$ such that
$b\neq c$. It means, that if player don't change his/her first choice, then he/she always win a prize.

Similar case in quantum Monty Hall game was considered by Flitney and Abbott in \cite{flitney2002quantum}, where the authors assumed that the players are permitted to select quantum strategies. As a result, if both players have access to quantum strategies, then there is a Nash equilibrium amongst mixed quantum strategies. We will show that there exist initial quantum state $\rho_{AB}$ for which the game is fair without permission to quantum strategies of the players.

Now consider another quantum state
\begin{align}
\hat \rho_{AB} &= \frac{1}{6} (\ket{01}+\ket{10})(\bra{01}+\bra{10}) \nonumber\\
&  +\frac{1}{6} (\ket{02}+\ket{20})(\bra{02}+\bra{20}) \\
&  +\frac{1}{6} (\ket{12}+\ket{21})(\bra{12}+\bra{21}). \nonumber
\end{align}
Here we achieved opposite results. For any 
$a,b,c\in\{0,1,2\}$ such that $ a\neq b$, $b\neq c$ and $a\neq c$ we have $\bra{a} 
\hat\rho_A^{B= b,C= c} \ket{a}=1$. Hence, in this case, if the player behaves like in classical scenario, then
his/her probability of winning is equal to $1$.

\begin{figure}
\centering
\includegraphics[scale=0.4]{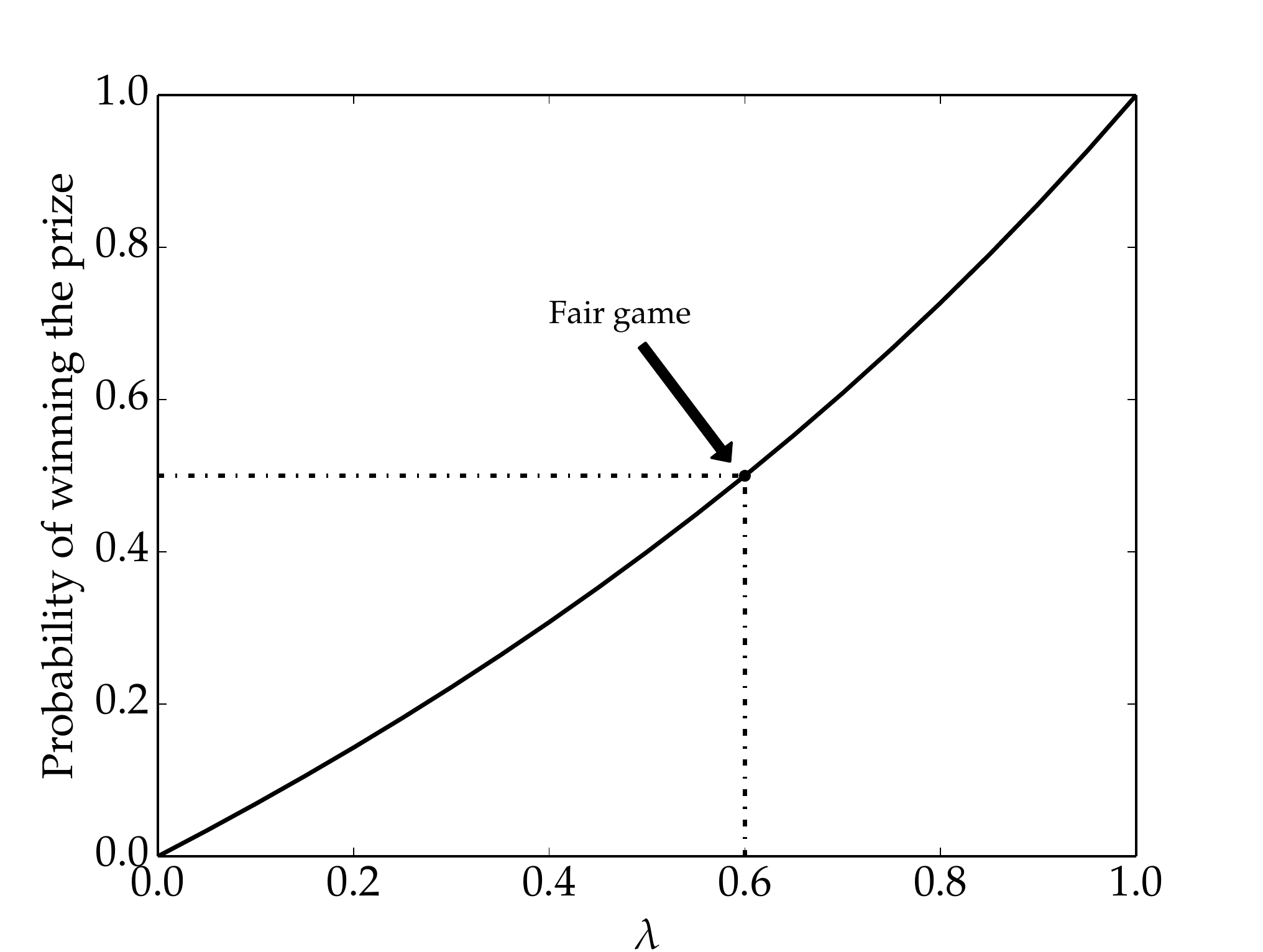}
\caption{Probability of winning the prize, when player does not change the door for the initial state given by $\bar\rho_{AB} = \lambda \tilde\rho_{AB} + 
(1-\lambda)\hat \rho_{AB}.$}
\label{glosKurzyk:probMix}
\end{figure}

Subsequently, let us consider convex combination of $\tilde{\rho}_{AB}$ and $\hat{\rho}_{AB}$
\begin{equation}
\bar\rho_{AB} = \lambda \tilde\rho_{AB} + (1-\lambda)\hat \rho_{AB}.
\end{equation}
For all $\lambda \in [0,1]$ the state $\rho_{AB}$ satisfies conditions specified at the beginning of this section. 
The probability of winning the prize by the player, under assumption that the player's first decision remains unchanged, is shown in
Fig.~\ref{glosKurzyk:probMix}. Moreover, for $\lambda =0.6$ and $a,b,c\in\{0,1,2\}$, where $ a\neq b$ and $b\neq c$  we get $\bra{a}\bar \rho_A^{B= b,C= c} \ket{a}=\frac12$. Thus, the player and the host win with equal 
probability and the game is fair.

Last example is very unintuitive. Consider following state
\begin{align}
\breve \rho_{AB} &= \frac{1}{6} (\ket{00}+\ket{01})(\bra{00}+\bra{01}) \nonumber\\
&  +\frac{1}{6} (\ket{11}+\ket{12})(\bra{11}+\bra{12}) \\
&  +\frac{1}{6} (\ket{22}+\ket{20})(\bra{22}+\bra{20}) .\nonumber
\end{align}
As a result we have
\begin{align}
\!\!\breve\rho_{A}^{B=0,C=1} \!\!&=\!\! 
\begin{bmatrix}
\frac{1}{4} & 0 & 0 \\
0 & 0 & 0 \\
0 & 0 & \frac{3}{4}
\end{bmatrix}, & \!\! 
\!\!\breve \rho_{A}^{B=1,C=2} \!\!&= \!\!
\begin{bmatrix}
\frac{3}{4} & 0 & 0 \\
0 & \frac{1}{4} & 0 \\
0 & 0 & 0
\end{bmatrix}, & 
\!\!\breve \rho_{A}^{B=2,C=0} \!\!&= \!\!
\begin{bmatrix}
0 & 0 & 0 \\
0 & \frac{3}{4} & 0 \\
0 & 0 & \frac{1}{4}
\end{bmatrix}, \\
% % % % % % % % % % % % %
\!\!\breve \rho_{A}^{B=0,C=2}\!\! &= \!\!
\begin{bmatrix}
1 & 0 & 0 \\
0 & 0 & 0 \\
0 & 0 & 0
\end{bmatrix}, & 
\!\!\breve \rho_{A}^{B=1,C=0} \!\!&=\!\! 
\begin{bmatrix}
0 & 0 & 0 \\
0 & 1 & 0 \\
0 & 0 & 0
\end{bmatrix},&
\!\!\breve \rho_{A}^{B=2,C=1} \!\!&= \!\!
\begin{bmatrix}
0 & 0 & 0 \\
0 & 0 & 0 \\
0 & 0 & 1
\end{bmatrix}. \nonumber
\end{align}
Above results show that the player shouldn't change his first decision $b$ only if the $c=(b+2)(\!\!\mod 3)$-th door has been opened. However, the host can minimize the  player's chance to $\frac34$ by opening the $c=(b+1)(\!\!\mod 3)$-th door.

\section{Conclusions}
In this paper, we showed the approach  based on quantum inferring acausal structures to reasoning in quantum information theory. Considered structures can be identified with a generalization of Bayesian networks, which concept is based on investigations of Leifer and Poulin \cite{leiferpoulin2008}. Proposed methodology was used to construct quantum model inspired by scenario of Monty Hall game.

The model was investigated both in classical and quantum case, whereby we showed that the entanglement of quantum states has influence on results of reasoning. We show that there exist quantum state for which Monty Hall game is fair under assumption that the player and the host do not have access to quantum strategies.

\section*{Acknowledgements}
This work was supported by the Polish National Science Centre: D. Kurzyk under the research project number UMO-2013/11/N/ST6/03090, A. Glos under the research project number DEC-2011/03/D/ST6/00413.

%\bibliographystyle{plain}
%\bibliography{Quantum_inferring_acausal_structures_and_the_Monty_Hall_problem}   % 
%%name your BibTeX data base

\end{document}